# The quantum state of light in collective spontaneous emission

Offek Tziperman[1†], Gefen Baranes[2,3†], Alexey Gorlach[1†*], Ron Ruimy[1], Chen Mechel[1], Michael Faran[1,4], Nir Gutman[1], Andrea Pizzi[3] and Ido Kaminer[1*]

[1]Technion – Israel Institute of Technology, Haifa 32000, Israel
[2]Massachusetts Institute of Technology, Cambridge, Massachusetts 02139, USA
[3]Harvard University, Cambridge, Massachusetts 02138, USA
[4]Tel Aviv University, Ramat Aviv 69978, Israel

[†]equal contributors, [*]kaminer@technion.ac.il, [*]horlach@campus.technion.ac.il

**Collective spontaneous emission occurs when multiple quantum emitters decay into common radiation modes, resulting in enhanced or suppressed emission. Here, we find the quantum state of light collectively emitted from emitters exhibiting quantum correlations. We unveil under what conditions the quantum correlations are not lost during the emission but are instead transferred to the output light. Under these conditions, the inherent nonlinearity of the emitters can be tailored to create desired photonic states in the form of traveling single-mode pulses, such as Gottesman-Kitaev-Preskill and Schrödinger-cat states. To facilitate such predictions, our work reveals the multi-mode nature of collective spontaneous emission, capturing the role of the emitters' positions, losses, interactions, and beyond-Markov dynamics on the emitted quantum state of light. We present manifestations of these effects in different physical systems, with examples in cavity-QED, waveguide-QED, and atomic arrays. Our findings suggest new paths for creating and manipulating multi-photon quantum light for bosonic codes in continuous-variable-based quantum computation, communications, and sensing.**

**Introduction**

Finding an efficient way to create quantum-correlated states of photons is a long-standing open problem in quantum optics. Substantial efforts are invested toward efficiently generating such quantum light states in various regimes of the electromagnetic spectrum, which will allow for highly desired applications in quantum technologies. For example, quantum states of light that can encode quantum information serve as necessary resources for continuous-variable quantum information processing [1], a rising approach complementing the conventional qubit encoding of discrete-variable quantum information processing. Applications of specific quantum states include squeezed light for enhanced sensing in spectroscopy and metrology [2], as well as non-Gaussian states of light such as Schrödinger's cat and Gottesman-Kitaev-Preskill (GKP) states that can support intrinsic error correction and fault-tolerant continuous-variable quantum computation [3-6].

In the optical regime, pioneering works in this field demonstrated the creation of squeezed light [7-11], bright squeezed vacuum [12-15], Fock states [16], photon-added coherent states [17], Schrödinger kittens [18], and few-photon cat states [19]. In the microwave regime, multi-photon Schrödinger cat [30,31] and GKP [32,33] states have been created, but only in stationary modes. Despite this encouraging progress, mechanisms to create high-quality multi-photon quantum states in the optical regime are scarce. In addition, it is desirable to create such states in the form of traveling pulses, rather than stationary modes, for useages in quantum communications and quantum networks. Creating such traveling pulses of continuous variable states is still a challenge across the electromagnetic spectrum [88]. The development of new mechanisms is consequently an active research avenue [20-22].

Creation of the desired non-Gaussian light states requires nonlinearity. In the optical range, the most common forms of nonlinearities are the natural $\chi^{(2)}$ and $\chi^{(3)}$ of transparent optical/bulk materials [23-24]. However, these nonlinearities are relatively weak and inefficient. Despite encouraging progress, they currently do not reach the single photon level in the low loss environments required for creating non-Gaussian states of light [90]. Alternatively, one can draw the nonlinearity from quantum emitters, such as atoms, quantum dots, or superconducting qubits [25-26], whose interaction with light can be strongly nonlinear, as captured by the classic Jaynes-Cumming model [27] in cavity-QED [28,29]. This approach is very successful in the generation of single or few photons [19] from a single emitter. However, creating the desired multi-photon quantum states from a single emitter requires very strong coupling and long lifetimes, which only exist in circuit QED [30-33] in the microwave range, currently not applicable in the optical range.

Our work investigates a fundamentally different strategy for the creation of desired multi-photon states: using nonlinear correlated emitters. The most notable example of collective light emission from quantum systems is that of superradiance and subradiance, discovered by Dicke in 1954 and heavily studied since then [34]. In Dicke superradiance, excited emitters can spontaneously decay in a collective fashion, emitting a short and intense pulse. Conversely, in subradiance, the different paths for the emitters to decay destructively interfere, leading to a prolonged emission time. Recently, superradiance and subradiance have attracted a resurgence of interest, including theoretical studies of superradiance in atomic arrays [35-40] either in free space or coupled to waveguides, experimental studies with cold atoms [41-44] or quantum dots [45-48], and recent studies that showed control of both superradiant and subradiant states [49,50].

Historically, most works on superradiance focused on the dynamics of the emitters, rather than on the quantum state of the emitted light itself. Still, some notable exceptions did look at the quantum properties of the radiation, such as the works by Bonifacio *et al*. [51,52] that investigated the quantum fluctuations of the light in the cavity as a function of time. More recently, the topic has seen a spike in interest, with works that have suggested superradiant states for quantum sensing [53], analyzed the multimode structure of superradiance [54], and showed a class of entangled superradiant states that emit classical light [55]. In addition, a few previous works have developed methods to create entangled states of emitters that will emit quantum light [56,57]. These have, however, considered only the low excitation "linear" regime, for which the number of excitations is small compared to the number of emitters, implying a simple one-to-one mapping between the emitting system and the emitted state of light [58]. No work so far has provided the full quantum optical picture of collective spontaneous emission, including the high excitation "nonlinear" regime. Indeed, many open questions remain on how to utilize collective spontaneous emission from correlated emitters to create desirable quantum states of light.

Here we show that the intrinsic nonlinearities emerging in quantum-correlated systems have a profound effect on the photonic states created in their collective spontaneous emission. Specifically, when correlated emitters undergo spontaneous emission in the nonlinear regime – involving high excitations of their collective states – they induce distinct quantum properties on the emitted photons. We show specific conditions under which super- and subradiance exploit this intrinsic nonlinearity to create desirable quantum states of light such as GKP and Schrödinger-cat states.

Quantum correlations are generally expected to be diminished by the nonlinearity of collective spontaneous emission, since it makes the pulse shape depend on its photon number state content.

Nevertheless, we find that for a surprisingly wide range of parameters, collective spontaneous emission creates an almost pure single-mode photonic state. This process then imprints the information of the correlated emitters' state onto the output radiation, offering a path to quantum states of light with high fidelity. As examples, we consider collections of emitters coupled to a waveguide or placed in a cavity that itself couples out to the continuum. We investigate the effect of emitter positions and interactions on the superradiant state transfer and find at which spacings superradiance prevails. We also investigate the quantum state of light in subradiance, creating different forms of quantum light such as Schrödinger cat states even from simple uncorrelated initial conditions. In all cases, we investigate both the linear and nonlinear regimes of collective spontaneous emission and derive the Wigner functions of the emitted states.

These findings establish the phenomenon of collective spontaneous emission as a promising tool for controllably translating correlated states of matter to quantum states of light, even deep into the nonlinear regime. The nonlinear regimes are found to be particularly useful for the creation of desirable quantum light states using a modest number of emitters. Our work contributes to pioneering efforts in the emerging field of many-body quantum optics, where light-matter interactions of many-body systems provide platforms for creation, storage, and unitary operations on continuous-variable quantum states of light and matter.

## Results

Our main goal is to find the quantum state of the emitted light in collective spontaneous emission. We describe the light state, as well as the state of the other involved subsystems (for example, of emitters and cavity field) in terms of Wigner quasi-probability functions, each carrying the entire quantum state information of a certain Hilbert space [59,60]. We link the output radiation to the initial state of the emitters and unveil under which conditions the Wigner function of the radiation contains negative values, which are regarded as evidence for non-classical optical properties [61].

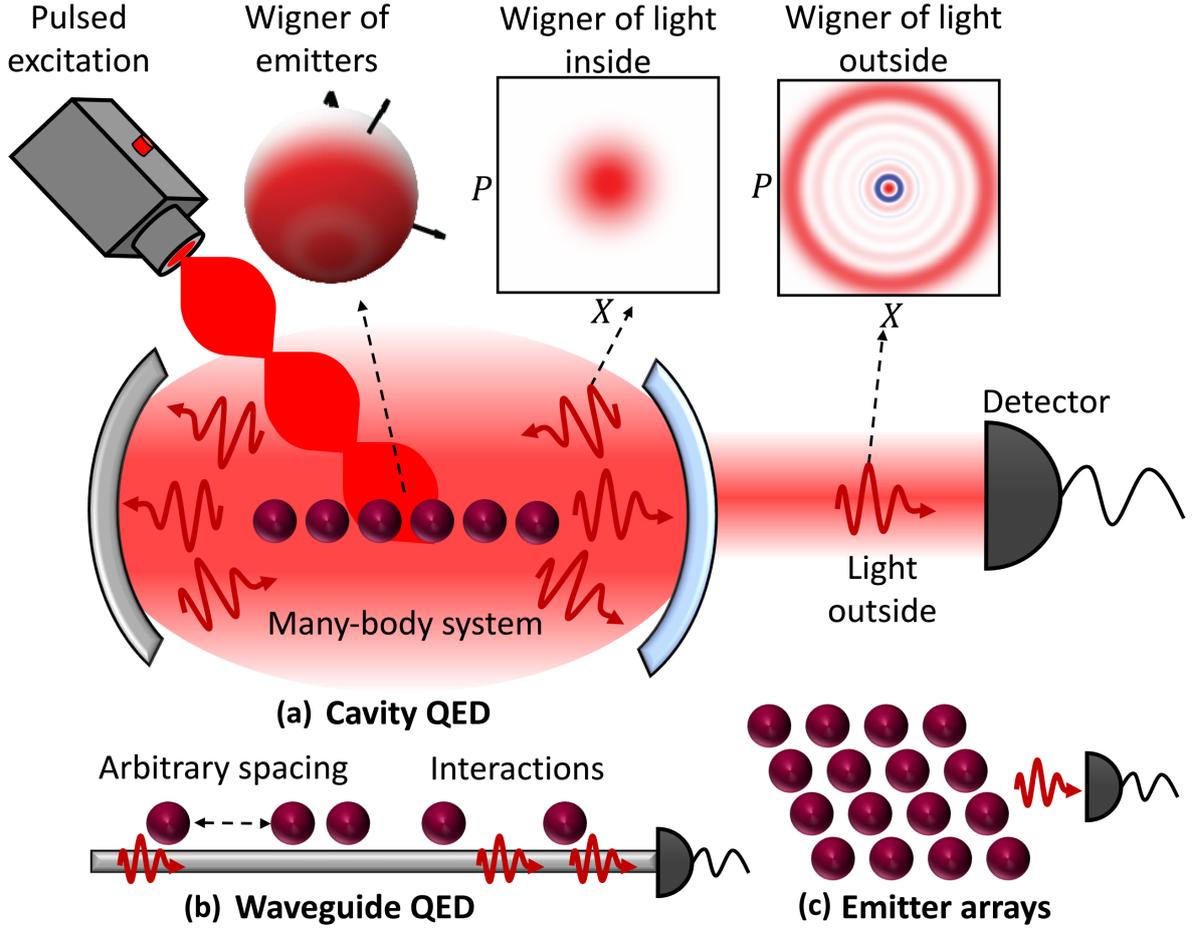

**Figure 1: The quantum state of light emitted in super- and subradiance.** We describe the interaction between a quantum-correlated system of emitters, the quantized electromagnetic field of a waveguide or a cavity, and the continuum of free-space modes outside of it. We exemplify the theoretical framework in three specific scenarios. **(a)** Cavity QED: Coherently excited the emitters radiate into a cavity mode, which gradually leaks outside through a semitransparent mirror (on the right) and is collected by a detector. **(b)** Waveguide QED: The excited emitters are arranged in one dimension and are coupled to a waveguide. **(c)** Atomic arrays: The emitters are positioned in two dimensional arrays in free space or in a cavity. In all three scenarios, we describe the quantum state of the pulse emitted from the system into a waveguide or to free space.

### 1. Theory of spontaneous emission from correlated-emitters

Consider $N$ emitters at positions $\{r_n\}_{n=1}^N$. The general Hamiltonian of the entire system reads:

$$H = H_S + H_{SF} + H_F, \tag{1}$$

where $H_S$ is the emitting system Hamiltonian, containing the emitters' energy levels and their internal interactions, $H_{SF}$ is the interaction term between the system and the field, containing coupling to the continuum of modes, and $H_F$ is the free-field Hamiltonian.

We first consider the case of cavity QED, extending to other platforms later. In this case, the system Hamiltonian is $H_S = H_E + H_{EC} + H_C$, where $H_E = \Omega_0 \sum_{n=1}^{N} \sigma_n^z = \Omega_0 S^z$ describes the energy of the emitters, treated as two-level systems with transition frequency $\Omega_0$ (henceforth, $\hbar = 1$). The term $H_{EC} = \sum_{n=1}^{N} g(\mathbf{r}_n) c^\dagger \sigma_n^- + g^*(\mathbf{r}_n) c \sigma_n^+$ describes the interaction between the emitters and the cavity mode in the rotating wave approximation, with creation (annihilation) operators $c^\dagger$ ($c$), spin ladder operators $\sigma_n^\pm = \sigma_n^x \pm i \sigma_n^y$, and $g(\mathbf{r}_n)$ the coupling strength at position $\mathbf{r}_n$. The cavity Hamiltonian reads $H_C = \Omega_1 c^\dagger c$. The interaction Hamiltonian $H_{SF} = \sqrt{\kappa/2\pi} \int_0^\infty (c^\dagger a_\omega + c a_\omega^\dagger) d\omega$ describes the coupling of the cavity field to the light outside of it (e.g., the free-space continuum or a waveguide). The operators $(a_\omega, a_\omega^\dagger)$ annihilate and create photons of frequency $\omega$ in the field outside the cavity, respectively. The decay rate $\kappa$ is assumed independent of frequency due to the Markov approximation for the coupling to the outside field [62]. Finally, $H_F = \int_0^\infty \omega a_\omega^\dagger a_\omega d\omega$ is the Hamiltonian of the light outside of the cavity, (e.g., of a one-dimensional free-space continuum or in a waveguide).

We summarize the theory in three steps: (i) Derive the full internal dynamics of the system, here the emitters and of the cavity field, while tracing out the continuum of modes. (ii) Find, outside of the cavity, the temporal shape of the mode(s) into which most of the light is emitted. (iii) Use this information to integrate the combined system and spatiotemporal modes of interest and obtain the quantum state of the pulse emitted from the system. The final step is made possible by adapting the recently developed input-output formalism with quantum pulses [63,64].

Step (i): Solve the internal dynamics of the system (emitters and cavity field). We trace out the light leaking outside of the cavity and obtain a Lindblad equation for the density matrix $\rho_S$ of the inner system "cavity field + emitters":

$$\dot{\rho}_S = \mathcal{L}[\rho_S] = i[\rho_S, H_S] + \mathcal{D}[\rho_S], \tag{2}$$

with Hamiltonian $H_S = H_E + H_C + H_{EC}$ and dissipator $\mathcal{D}[\rho_S] = \kappa c \rho_S c^\dagger - \frac{\kappa}{2}\{c^\dagger c, \rho_S\}$. The former is composed of terms describing the emitters, cavity field, and interaction between them, respectively, whereas the latter accounts for leakage from the cavity at a rate $\kappa$. We solve Eq. (2) numerically without making further approximations. Note that, unlike the case of emission in free space, in which light is forever "lost" and can be treated within a Markov approximation, here we do not use this approximation. This treatment is especially important when the cavity loss rate $\kappa$ is not too large, and the light in the cavity can be reabsorbed by the emitters. From the point of view of the emitters, the resulting Rabi-type oscillations is non-

Markovian, having energy flowing back and forth between emitters and cavity. From the solution $\rho_S(t)$ of Eq. (2), we extract the reduced density matrices $\rho_{\text{emitters}}(t)$ and $\rho_{\text{cavity}}(t)$ and compute the respective time-dependent Wigner functions. In particular, the cavity field begins and ends in a vacuum state, $\rho_{\text{cavity}}(t = 0) = \rho_{\text{cavity}}(t = \infty) = |0\rangle\langle 0|$, and the emitters end in their ground state, $\rho_{\text{emitters}}(t = \infty) = |0\rangle\langle 0|$.

Step (ii): Identify the emission wavepacket mode(s). To reconstruct the field outside the cavity, we first identify the most occupied modes of the output radiation. To this end, we compute the first-order correlation function using the input-output relation: $\Gamma^{(1)}(t_1, t_2) = \kappa\langle c^\dagger(t_1)c(t_2)\rangle = \langle a^\dagger(t_1)a(t_2)\rangle$ [65,66], where $a(t) = 1/\sqrt{2\pi} \int_{-\infty}^{\infty} d\omega\, a_\omega e^{i\omega t}$ is the Fourier transform of $a_\omega$. Next, we diagonalize $\Gamma^{(1)}(t_1, t_2) = \sum_i n_i \eta_i^*(t_1)\eta_i(t_2)$ [67], such that each *principal mode $i$* is characterized by a complex temporal shape $\eta_i(t)$ and photon occupancy $n_i$. Sorting the principal modes in descending order by their photon occupancy (mean number of photons), we choose a minimal set of modes that carry all the information of the emission. Each mode $i$ is associated with $a_i^\dagger = \int_0^\infty \eta_i(t')a^\dagger(t')dt'$, with $[a_i, a_j^\dagger] = \delta_{ij}$. The numbers $n_i$ correspond to the mean photon numbers in each mode.

We are looking for scenarios where a single mode of light has quantum correlations, after the others are traced out. For this purpose it is enough to calculate the quantum state in mode $i = 0$, while tracing out the others as loss. We tag such a *dominant mode* by $a_0^\dagger$:

$$a_0^\dagger = \int_0^\infty \eta_0(t')a^\dagger(t')dt'. \tag{3}$$

Step (iii): Solve the dynamics of the dominant modes. Next, we adapt the recently developed input-output theory with quantum pulses [63,64,67] and apply it for the first time to a quantum many-body emitters system. With this theory, we solve the internal dynamics again, but this time include explicitly the dominant emitted mode $a_0^\dagger$. The master equation that we numerically integrate is thus:

$$\dot\rho = i[\rho, H_S + H_I] + \sum_i \left(L_i \rho L_i^\dagger - \frac{1}{2}\{L_i^\dagger L_i, \rho\}\right), \tag{4}$$

where $H_I(t) = \frac{i}{2}\sqrt{\kappa}\left(g_0^*(t)c^\dagger a_0 - g_0(t)c a_0^\dagger\right)$ is the interaction between the system and the dominant mode, $L_0(t) = \sqrt{\kappa}c + g_0^*(t)a_0$, $g_0(t) = -\eta_0^*(t)/\sqrt{\int_0^t |\eta_0(t')|^2 dt'}$, and $L_i$ are the jump operators for possible additional decay channels (e.g., to modes transversal to the cavity axis), or dephasing channels. The combined density matrix $\rho$ includes the emitters, cavity field, and dominant temporal mode(s) of the propagating light pulse outside of the cavity. Here lies a key advantage of our approach – it reduces the Hilbert space to a finite size, no longer requiring an infinite number of photonic modes (as the emission into other temporal modes is taken into account and traced out).

We refer the reader to the Supplementary Material (SM) for the derivation of Eq. (4) (SM Sections II, III) using elements from [64,65]. Furthermore, we generalize the analysis for the cases of a one-dimensional waveguide (waveguide QED, SM Section IV), and three-dimensional free space (emitter arrays, SM Section V), where we also extend the theory from [64,65] to multiple spatiotemporal modes. In a similar manner, the theory can be applied to emissions into arbitrary photonic environments [68].

## 2. Quantum light emission in Dicke superradiance

To gain intuition into the problem, let us begin by studying the simplest situation of Dicke superradiance in a cavity. In this scenario, all the emitters are identical, initially excited and couple to the field in the same way ($g(\boldsymbol{r}_n) = g$), with only a single decay channel to the outside continuum. Sections 3-5 move beyond this scenario by considering different initial conditions, the emitters' positions, and further unwanted decay channels. We implement the theory above and describe light outside the cavity for initially fully inverted (i.e., excited) emitters, as can be prepared by a $\pi$-pulse. We then quantify what part of the initial emitters quantum state is lost during spontaneous emission due to the multimode nature of the emitted light, rather than being transferred to the emitted photons.

The dynamics of the system is shown in Fig. 2a, comprising the Wigner functions of the emitters (1st row, on the Bloch sphere) [59], of the cavity field (2nd row), and of the propagating dominant mode $a_0$ (3rd row). Energy is exchanged between the emitters and the cavity with Rabi-like oscillations until all the energy leaks out to free space, leaving the emitters in their ground state and the cavity in the vacuum state (final time in Fig. 2). Due to its entanglement to the emitters, the cavity field's Wigner function is always positive, whereas the outside light dynamically builds up negative parts in its Wigner function (as shown at time $t = 1.4g^{-1}$ in Fig. 2a). Furthermore, the final Fock-like state of the dominant mode resembles that is different from the initial emitters' Dicke state (a Fock state in the emitters), as quantified by a fidelity of only 71%. The deviations from a perfect transfer are due to the population of multiple temporal light modes due to the nonlinearity of collective spontaneous emission.

By making the master equation dimensionless, one realizes that, in the absence of emission into other spatial modes, the dynamics of the system is controlled by a dimensionless parameter, which we call *reabsorption efficiency* $\xi$ that reads:

$$\xi = \frac{2g^2 N}{\kappa^2}, \tag{5}$$

and quantifies how well the light emitted into the cavity can be reabsorbed by the emitters. A larger $\xi$ makes the system more nonlinear and the emitted light multimode (as will be shown below). The regime of $\xi > 1$ requires a strong light-matter coupling $g$ and a small leakage rate $\kappa$, leading to an efficient back-and-forth Rabi-type exchange of energy between the emitters and the cavity, as witnessed for instance, by the intensity of the emission outside the cavity $\langle a^\dagger a \rangle(t)$ (Fig. 2b). The parameter $\xi$ also controls the quantum state of the cavity field, the quantum state of the dominant mode of the emitted field outside the cavity, and the spectrum of this dominant mode (SM, Figs. S1, S2, S4).

Fig. 2c shows the entanglement entropy of the light "outside" with the system "inside" (emitters + cavity). At initial time $t = 0$, the outside light is in a vacuum state and is thus disentangled from the "inside" (Fig 2c). Entanglement then dynamically builds up during the emission. After peaking, the entanglement entropy decreases till vanishing corresponding to an empty cavity and emitters in the ground state at long times. As expected, large amounts of entanglement render each subsystem more mixed, and their Wigner functions positive. Indeed, a negative Wigner function (of the output radiation) is only possible at long times when entanglement ceases. Further characteristics of the light in the cavity and the modes of the emission are calculated in the SM.

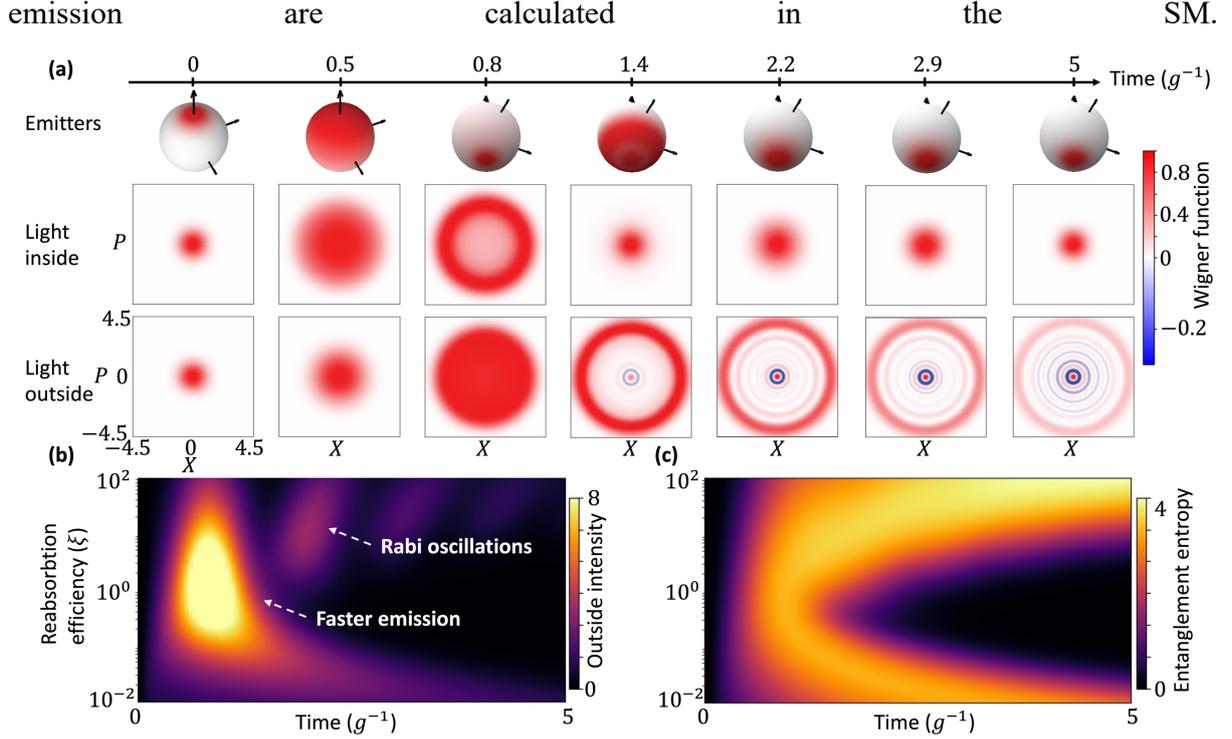

**Figure 2. Quantum state dynamics in Dicke superradiance.** (**a**) Wigner functions of the emitters (top), cavity field (middle), and propagating dominant mode outside of the cavity (bottom), at selected times. All Wigner functions are normalized by their maximum value, and the reabsorption efficiency is set to $\xi = 3$. (**b,c**) The intensity of the propagating dominant mode outside of the cavity (in units of the coupling $g$) and its entanglement entropy with the system "cavity + emitters", versus time and reabsorption efficiency $\xi$. Here, we consider $N = 10$ initially fully inverted emitters.

## 3. Shaping non-Gaussian states of the emission

We have shown that the spontaneous emission from fully excited emitters consists of nearly a single mode propagating light pulse with quantum features (Fig. 2c). A natural question to ask is how the output quantum state of light depends on the specific initial condition of the emitters.

Previous works have shown that in the linear regime, where the number of excitations is far smaller than the number of emitters, the quantum state of the emitters can collectively transfer to light [57-58]. However, most current platforms, such as atom arrays and superconducting qubits coupled to transmission lines, have only a modest number of strongly coupled qubits. For this reason, it is important to study also the strongly nonlinear regime of collective spontaneous emission.

We now consider superradiance without a cavity, so the emitters are coupled directly to the continuum of photonic modes. This scenario can be described by the formalism above by replacing the photonic operators of the cavity field $c$ by the collective lowering operator $S_-$ in Eq. (4) and in all the elements within. This way, the interaction Hamiltonian describes the direct coupling between the emitters and the continuum of modes or the dominant modes $a_0$. This description captures a range of systems including emitters in free space with a spatial extent much smaller than the emitted wavelength and emitters coupled to a one-dimensional waveguide with wavelength spacing.

In Fig. 3, we find that desired quantum states can collectively transfer from matter to light even in the nonlinear regime – i.e., when the number of excitations is close to the number of emitters, so the collective states are highly excited. For example, when the emitters are fully inverted by a $\pi$-pulse, the resulting photonic state resemble a propagating Fock state pulse (83% fidelity, in the simulation shown in Fig. 3a). The deviation from a perfect Fock state is caused by the multimode nature of the emission, induced by the strong nonlinearity of the fully inverted emitters. Collective Schrodinger's cat and GKP states of emitters transfer to Schrodinger cat and GKP states of light (with 99% and 97% fidelity in the simulations shown in Figs. 3b,c). In a follow-up paper [89], we show how to use the simplest form of collective nonlinearity of quantum emitters, spin squeezing, to entangle them into collective quantum states such as Schrodinger cat and GKP on the sphere. Furthermore, using known methods involving either cavity feedback, two-colored excitation, or post-selection, it is possible to create a spin-squeezed state of the emitters [76–78], resulting in the emission of squeezed non-Gaussian light (Fig. 3d). We emphasize that although squeezed light states are Gaussian, the light that is emitted from squeezed atomic states is non-Gaussian, due to the curvature of the Bloch sphere. Section VI of the SM provides details about the emitters' initial preparation.

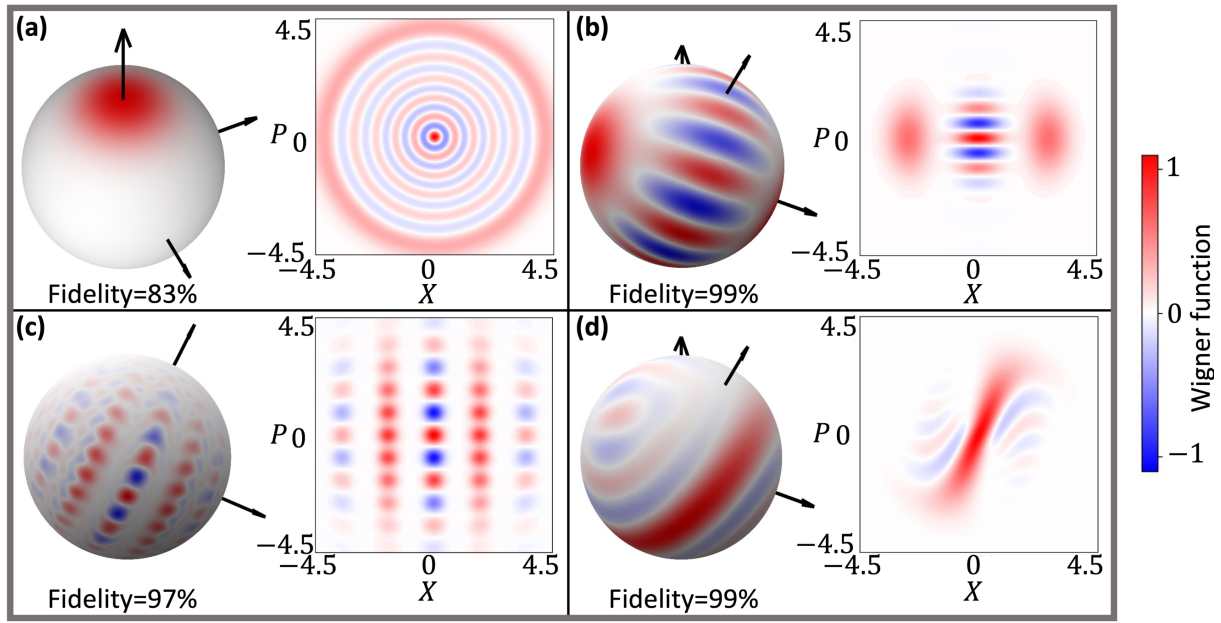

**Figure 3. Strongly nonlinear collective spontaneous emission transferring of quantum states from emitters to light**. Preparing correlated states of the emitters via a combination of coherent control pulses and nonlinearities, it is possible to obtain desirable output quantum states of light. The state of emitters and output light is shown in terms of their Wigner functions. **(a)** The emission from fully inverted emitters generates a Fock state with fidelity 83%. **(b,c)** Using second-order nonlinear interactions and coherent excitations of the superradiant system, it is possible to make it emit Schrodinger cat and GKP states with fidelities of 97% and 99%. **(d)** Emitters prepared in an atomic squeezed state result in the emission of a squeezed, non-Gaussian state of light. Here, we consider $N = 10$ emitters in panels **(a,b,d)** and $N = 40$ emitters in panel **(c)**, with no additional losses.

## 4. Beyond Dicke superradiance – varying emitters' positions and interactions

We now generalize our results beyond the subspace of symmetric Dicke states and study how the spatial arrangement of the emitters affects the output radiation. Our model is inspired by recent experimental breakthroughs that have demonstrated accurate control over the emitters' position [82,91], paving the way to engineering their light-matter couplings $g(\mathbf{r}_n)$ and the interactions between emitters. It is therefore important to study the effect of the emitters' positions and their interactions on the quantum state of the spontaneously emitted light. The steps (i-iii) of the recipe above can be repeated, but now on the full exponentially large Hilbert space of $N$ two-level emitters. In the absence of the symmetry under any emitters' permutation, destructive interference can lead to subradiance and dark states, hindering the spontaneous decay of the emitters into the waveguide or cavity. Instead, these states will eventually decay into other spatial modes outside of the waveguide or cavity.

To exemplify the effect of emitter positions, we first investigate the simple case of emitters coupled to a 1D waveguide (Fig. 1b). To ensure light emission in a single direction, we consider a mirror placed at $z_{\text{mirror}} = -\lambda_0/4$. The emitter position $z_n$ along the waveguide axis affect the light-matter couplings as $g(z_n) = g(0)\cos\left(\frac{2\pi z_n}{\lambda_0}\right)$ with $\lambda_0 = 2\pi c/\Omega_0$ being the wavelength at the frequency $\Omega_0$ and $c$ being the speed of light in the medium. Under a Markov approximation, we can account for the positions in the dissipative interactions of the emitters by redefining $S^\pm$ as $S^\pm = \sum_n \sigma_n^\pm \cos\left(\frac{2\pi z_n}{\lambda_0}\right)$. We include the waveguide-mediated interaction by $H_{\text{WMI}} = -\frac{i\Gamma_{1D}}{4}\left(\sum_{m\neq n}\left(e^{\frac{i2\pi|z_m-z_n|}{\lambda_0}} - e^{\frac{i2\pi(z_m+z_n+2z_{\text{mirror}})}{\lambda_0}}\right)\sigma_m^+\sigma_n^- - \text{h.c.}\right)$. The Hamiltonian of the system is then $H_S = \Omega_0 S^z + H_{\text{WMI}}$, for the free field is $H_F = \int_0^\infty \omega a_\omega^\dagger a_\omega d\omega$ and the Hamiltonian describing the interaction with the dominant mode is $H_I = \frac{i}{2}\sqrt{\Gamma_{1D}}(g_0^* S^+ a_0 - g_0 S^- a_0^\dagger)$ (SM Section IV for details) [83]. $\Gamma_{1D}$ is the single-atom decay rate into the waveguide.

Fig. 4a presents the Wigner functions of the light emitted from a 1D array of emitters coupled to a waveguide (as in Fig. 1b) as a function of the spacing between the emitters, for initial excitations of coherent, cat, half-excited Dicke, and fully excited states. In the special case of emitter spacings that are multiples of the transition wavelength, we get $H_{\text{WMI}} = 0$, and the system retrieves the Dicke limit (left column of Fig. 4a). For spacings that are not exactly multiples of the wavelength, but close to it, emission can still be very close to that predicted in the Dicke limit (compare first and second columns of Fig. 4a). By contrast, for spacings far from integer multiples of the wavelength, the state of the emitted light deviates substantially from that of the Dicke limit (center and left columns of Fig. 4a). This deviation can occur because of two different reasons: (i) $H_{\text{WMI}} \neq 0$, and the nonlinear interactions of the emitters create multimode entangled light emission, hindering the purity of the dominant mode when tracing out the others, e.g. $d = \lambda_0/4$ in Fig 4a. (ii) $H_{\text{WMI}} = 0$, but the emission is subradiant due to destructive interference of the radiation from different atoms, e.g. $\lambda_0/2$ in Fig. 4a.

To investigate whether the emission can result in pure single-mode quantum state far from the Dicke superradiant regime, we show the purity of the dominant photonic state in Fig. 4c. When the spacings are close to integer multiples of $\lambda_0$, energy is efficiently transferred to the emission (Fig. 4b). Then, the emitted state of light is almost a pure single-mode state, and the state of the emitters is mapped to that mode with high fidelity. In contrast, if the spacings are far from integer multiples of $\lambda_0$, the mode's purity drops. This reduced purity is either (i) due to the emission of a multimode entangled state, where each mode becomes mixed when the others are traced-out, or (ii) because the correlated emitters fail to reach the ground state, getting stuck in subradiant dark states, thus remaining entangled with the emitted light and reducing its purity.

An intriguing exception occurs for $\lambda_0/2$ spacing, featuring relatively pure subradiant emission (Figs. 4b, 4c). Interestingly, in this case, the Wigner function of the emitted light has negative parts (Fig. 4a, last column) and resembles a Schrodinger cat state, even when the emitters are initially in a coherent state. These results suggest that placing the emitters at $\lambda_0/2$ spacing can realize quantum light emission even from uncorrelated initial conditions that are more straightforward to prepare, for example, by coherent control with a $\pi/2$-pulse incident perpendicular to the waveguide.

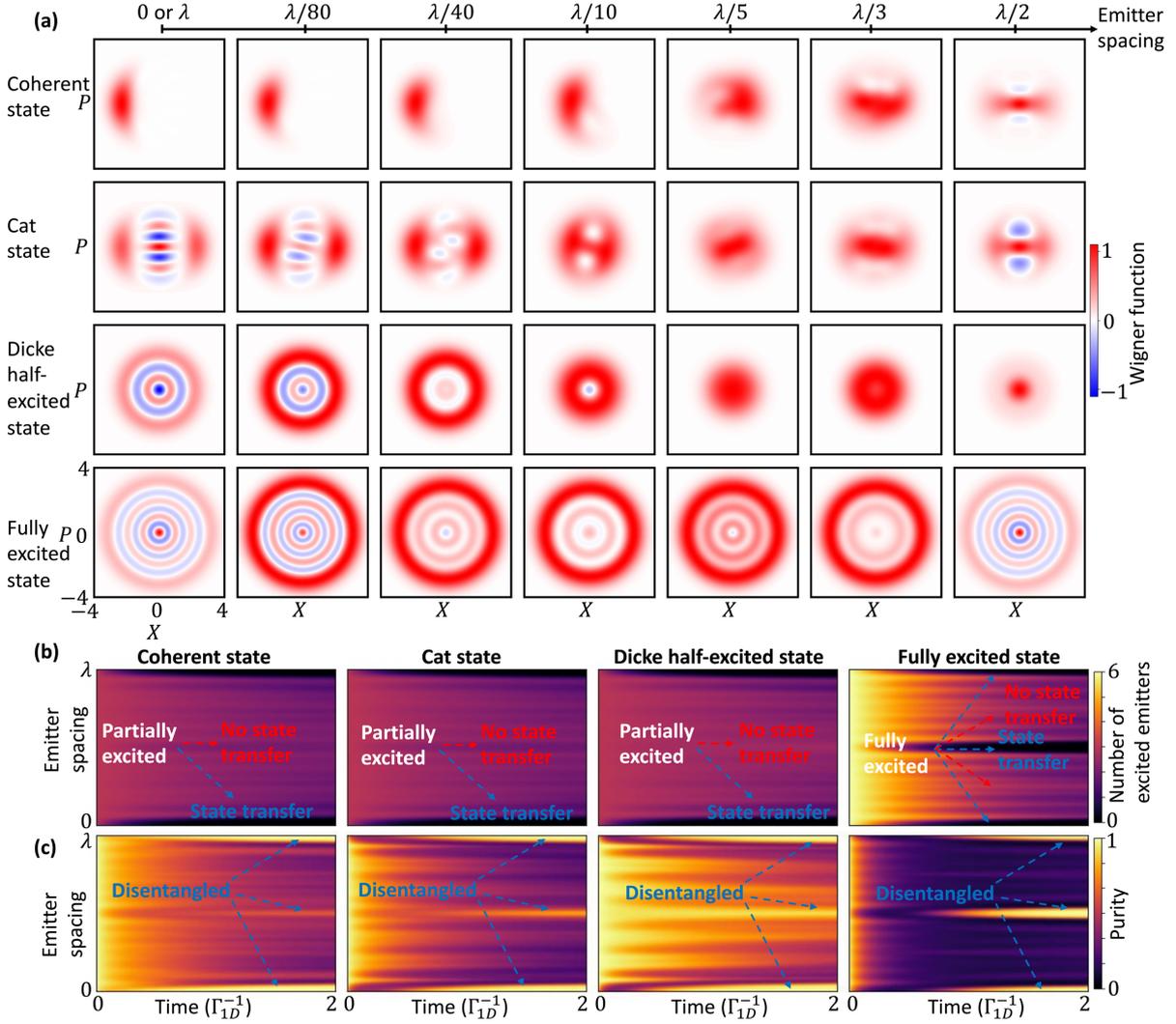

**Figure 4. Shaping the quantum state of collective spontaneous emission through emitters' positioning.** The quantum state of light emitted in waveguide QED for various spacings and initial conditions of the emitters. (**a**) Wigner function of the light emitted into a waveguide for different initial conditions (coherent state, cat state, Dicke half-excited state, and fully excited state). (**b**) Inversion population (i.e., number of excited emitters) versus spacing and time. We observe that for spacing far from a multiple of the wavelength $\lambda_0$, subradiance hinders the decay of the emitters to their ground state. (**c**) Purity of the atomic state. We observe that the emitted light remains disentangled from the emitters and from the other modes when the spacings are multiples of $\lambda_0/2$. Here, we considered $N = 6$ emitters.

## 5. Robustness of quantum light generation to decoherence

To assess the robustness of our findings to decoherence channels, we introduce these channels by the last term of Eq. (4), modeling spontaneous emission into undesired modes and dephasing of the emitters. The decoherence channels can be split between collective and independent decoherence channels. We account for the collective decoherence using two different factors: the first is decay to unwanted *spatial modes*, which we consider by adding the collective decay term $L^{\mathrm{col}} = \sqrt{\gamma'} S^-$, parameterized by the single emitter spontaneous emission rate into free space $\gamma'$. The second is decay into unwanted *temporal modes* because the emitted light cannot be described entirely as a single-mode state, as considered in Eq. (4) by the Lindblad term $L_0$.

The decay to unwanted spatial modes exhibits a collective nature for certain configurations of the emitters, yielding enhancement by superradiance or suppression by subradiance. For example, it was shown that certain configurations of the emitters can make the emission into the unwanted modes subradiant (thus, suppressed) while preserving superradiance in the wanted mode [79].

In Fig. 5a,b, we consider cavity QED emission hampered by a collective superradiant decay channel into unwanted spatial modes. Generally, the longer the emission duration, the lower the fidelity of the dominant mode (calculated relative to the initial state of the emitters), due to the action of the other decoherence channels. The fastest emission from the system requires both a fast decay of the emitters into the cavity (large $g^2/\kappa$), and a fast decay of the light out of the cavity (large $\kappa$). The highest fidelity is thus achieved for $\xi \approx 0.5$ (dashed white line in Fig. 5a,b), corresponding to the shortest emission time.

In addition to the collective decoherence channels, the emitters' also decohere through independent channels, which include inhomogeneous spectral broadening causing dephasing, and independent decay to undesired spatial modes. In Fig 5c,d, we consider waveguide QED hindered by different unwanted decoherence channels. Dephasing and independent decay are modeled by adding the terms $L_n^{\text{dep}} = \sqrt{\gamma'}\sigma_n^z$, or $L_n^{\text{ind}} = \sqrt{\gamma'}\sigma_n^-$ to the master equation, with $1 \leq n \leq N$ denoting the emitter index. For a quantitative comparison of the effect of each channel (collective superradiant decay $L^{\text{col}}$, independent decay $L_n^{\text{ind}}$, and dephasing $L_n^{\text{dep}}$), we parametrize all unwanted channels with the same decoherence rate $\gamma'$ and simulate the process under the effect of one channel at a time, while setting the others to zero.

We observe that the collective decay channel ruins the quantum state of the emitted light the most, as shown by the lowest fidelity for this channel in Fig 5c,d. Indeed, the quantum light is more robust to the effect of independent decoherence channels, which for example can be of the same order as the single emitter decay rate into the 1D waveguide $\Gamma_{1D}$, while still preserving the quantum nature of the emission (as seen by the high fidelities in Fig 5c,d for these values). Altogether, following the above considerations, if the number of emitters is increased, while keeping their total number of excitations constant, the fidelities can be improved. In this case, a larger fraction of light will be emitted into the wanted mode, since the superradiant decay rate increases with the number of emitters while the independent decay rate scales only as the number of excitations.

We also quantify the effect of undesired losses with respect to the initial condition of the emitters in all panels of Fig. 5. In general, both Fock and cat states are very sensitive to decoherence [81], as the loss of a single photon tends to wash away negative parts of the Wigner function. Indeed, the overall fidelity of the cat state (Fig. 5b,d) is larger than that of the Fock state (Fig. 5a,c). This difference can be attributed to the cat state having a lower initial excitation, and thus being less sensitive to photon loss.

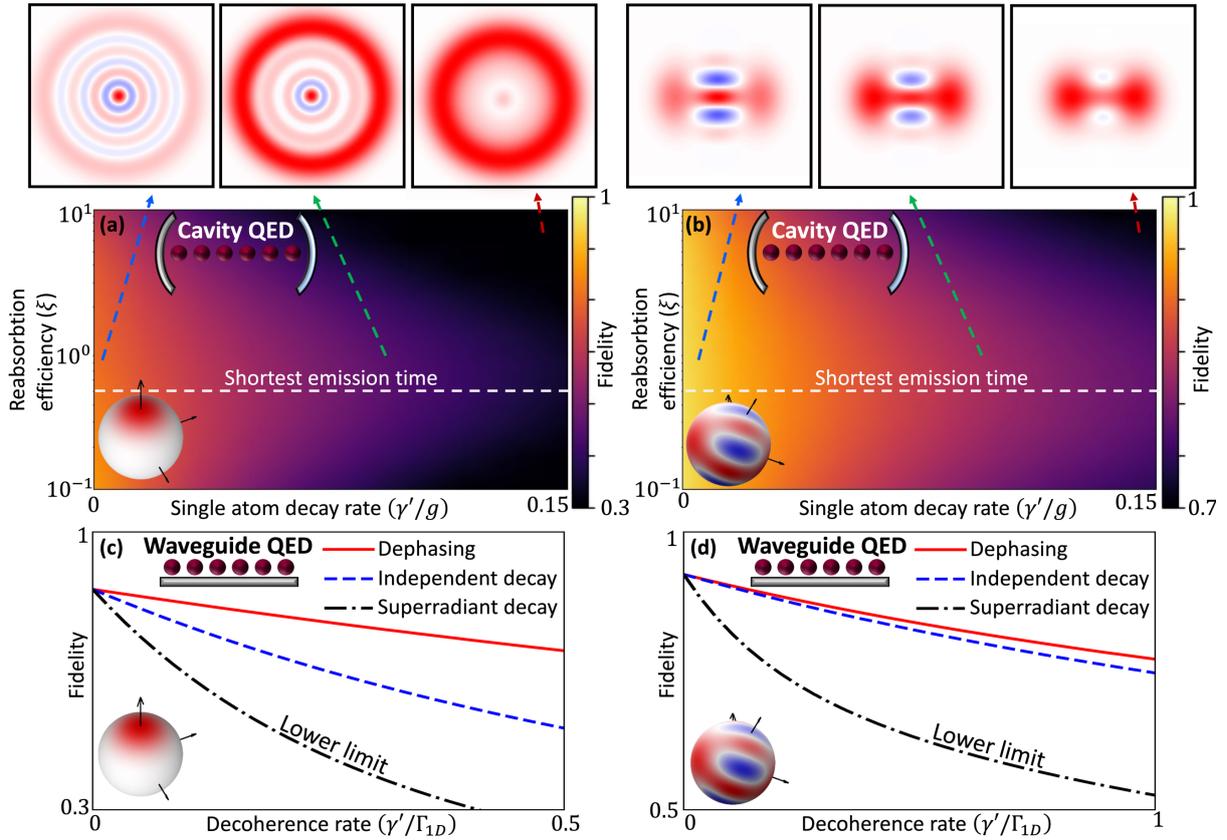

**Figure 5. Robustness of quantum light generation against spurious losses. Cavity QED:** Fidelity of the light emitted into the dominant propagating mode outside the cavity, calculated by considering the ideal state as the initial state of the emitters. The fidelity map is plotted as a function of the cavity decay rate and the single atom decay rate. The latter quantifies undesired spontaneous emission into spatial modes outside the cavity. The highest fidelities are reached for a reabsorption efficiency $\xi = 0.5$ (white dashed line). **Waveguide QED:** Emitters located at $z_n = n\lambda_0$ are coupled to a one-dimensional waveguide. We investigate the fidelity as a function of different decoherence rates, such as inhomogeneous dephasing as well as independent and collective decays. The initial conditions are the all excited state **(a,c)** and 2-legged cat state **(b,d)** respectively, and are plotted on the Bloch sphere as insets. Here, we considered $N = 6$ emitters.

**Discussion**

We studied the quantum state of light emitted in collective spontaneous emission and showed its strong dependence on the intrinsic nonlinearity of systems with quantum correlations in several platforms, both in the super- and subradiant regimes.

In the case of superradiance, we found that for weak interactions and near-identical emission from all the emitters, the emission consists of mostly a single dominant spatiotemporal mode. In this case, the initial state of the emitters are mapped to a traveling pulse of light with high fidelity. We showed the robustness of the state transfer by investigating the effects of losses, positions, interactions, and initial conditions of the emitters. In the case of subradiance, we found that controlling the emitter positions can cause even initially uncorrelated systems to emit quantum-correlated light similar to Schrödinger cat states.

Our work describes the emission process without neglecting the nonlinear corrections due to the curvature of the Bloch sphere, fully capturing the nonlinear regime of superradiance. Notable contributions [56], [57] regarded the linear regime, in which the Hilbert space of the emitters can be approximated as a harmonic oscillator [58]. In that regime, the density matrix of the emitted light is the same as the initial density matrix of the emitters. The main contrast between the linear and nonlinear regimes is that the curvature of the Bloch sphere enables to create arbitrary non-Gaussian single-mode states of light, even when considering only Gaussian spin operations (such as rotations and spin squeezing) [89]. Besides the nonlinearity, our work adds to previous investigations in the field the first multi-mode description of superradiant and subradiant emission for any arbitrary initial condition of emitters and arbitrary emitter positions.

Looking forward, our findings suggest a novel strategy for the creation of quantum states of light for photonic quantum applications: by manipulating the correlated quantum state of an ensemble of emitters in their nonlinear regime or outside the Dicke manifold. This strategy reduces the number of emitters needed to generate desired quantum photonic states, by relaxing the condition that the number of emitters be much larger than the number of excitations. Such a strategy also provides a novel form of nonlinearity, emerging directly from the quantum emitters or from their joint correlated states. Using the emitters' nonlinearity for creation of quantum states of light also benefits from the many recent advances in controlling the emitters [80,82,84–86].

Exciting prospects emerge when considering systems of emitters based on other areas of correlated many-body physics, as in certain solid-state systems. Each system will have intrinsically different correlations and interactions between the emitters, translating the correlated emitters' states into correlations in their emitted radiation, potentially involving entanglement across multiple radiation modes. In this way, many-body physics can transform into a platform to generate quantum light with not only rich photon statistics and coherence, but also rich entanglement structures such as cluster states and NOON states, which are highly desirable, especially in optical frequencies.

**Code Availability Statement**

The code supporting the findings of this study uses QuTiP [87] and is available at:

https://github.com/offektziperman/spontaneous-emission-from-correlated-emitters